\documentclass[prd,aps,preprint,tightenlines,nofootinbib,superscriptaddress]{revtex4}

\usepackage{amsmath}
\usepackage{epsfig}
\usepackage{axodraw}

\newcommand{\insertfig}[2]{\mbox{\epsfxsize=#1cm \epsfbox{#2.eps}}}
\def\Journal#1#2#3#4{{#1} {#2} (#4) #3 }
\def\Journal#1#2#3#4{{#1} {#2} (#4) #3 }

\def\PRL{\em Phys. Rev. Lett.}

\begin{document}

\title{\mbox{\hspace*{-0.35cm}Higgs-Higgsino-Gaugino Induced Two Loop Electric Dipole Moments}}

\author{Yingchuan Li}
\email{yli@physics.umd.edu} \affiliation{Department of Physics,
University of Wisconsin, Madison, Wisconsin 53706 USA}
\author{Stefano Profumo }
\email{profumo@scipp.ucsc.edu} \affiliation{Department of Physics
and Santa Cruz Institute for Particle Physics,
University of
California, 1156 High St., Santa Cruz, CA 95064, USA}
\author{Michael Ramsey-Musolf}
\email{mjrm@physics.wisc.edu} \affiliation{Department of Physics,
University of Wisconsin, Madison, Wisconsin 53706 USA}
\affiliation{Kellogg Radiation Laboratory, California Institute of Technology, Pasadena, CA 91125 USA}
\date{\today}

\begin{abstract}

\noindent We compute the complete set of Higgs-mediated chargino-neutralino
two-loop contributions to the electric dipole moments (EDMs) of the
electron and neutron  in the minimal supersymmetric standard model
(MSSM). We study the dependence of these contributions on the
parameters that govern CP-violation in the MSSM
gauge-gaugino-Higgs-Higgsino sector. We find that contributions
mediated by the exchange of $W H^\pm$ and $ZA^0$ pairs, where
$H^\pm$ and $A^0$ are the charged and CP-odd Higgs scalars,
respectively,  are comparable to or dominate over those mediated by the exchange of
neutral gauge bosons and CP-even Higgs scalars. We also emphasize that the result of this complete set
of diagrams is essential for the full quantitative study of a number
of phenomenological issues, such as electric dipole
moment searches and their implications for electroweak baryogenesis.
\end{abstract}

\maketitle

\section{introduction}

The search for CP-violation (CPV) beyond that of the Standard Model
(SM) lies at the forefront of nuclear and particle physics. Perhaps
the most powerful probes for new CPV are searches for permanent
electric dipole moments (EDMs) of the electron, neutron, and neutral
atoms. Null results obtained from these searches have placed
stringent constraints on CPV in the strong sector of the SM, while
present and expected future sensitivities lie several orders of
magnitude away from expectations based on CPV associated with the
phase of the Cabibbo-Kobayashi-Maskawa (CKM) matrix. Various
scenarios for CPV connected to new physics at or above the
electroweak scale naturally imply the existence of non-vanishing
EDMs that could be observed in future experiments. Thus, a
comprehensive program of EDM searches could uncover either CPV
associated with the \lq\lq $\theta$-term" of the QCD Lagrangian, new
electroweak scale physics, or both. Each possibility has potentially
significant consequences for cosmology. The  Peccei-Quinn mechanism
proposed to explain the vanishingly small value of $\bar\theta$
implies the existence of an axion that could account for the cold dark matter (CDM), while new electroweak scale CPV
could help in explaining the observed abundance of baryonic matter
through the mechanism of electroweak baryogenesis (EWB).

Among the most theoretically attractive possibilities for new physics
is supersymmetry (SUSY). SUSY provides an appealing solution to
the naturalness problem of the SM. However, SUSY
has to be softly broken to be consistent with experimental
observations. In order to solve the naturalness problem, the SUSY
breaking scale should be not much higher than a few TeV. While the
exact mechanism of soft SUSY breaking is not yet known, its effect is
encoded into the soft terms in the low-energy realization of this
scenario. In the minimal supersymmetric standard model (MSSM), the
presence of soft terms implies the existence of 40 additional
CPV phases beyond the single phase of the CKM matrix in the SM. As
there exists no known  {\em a priori} reason for these phases to be
suppressed, one expects rather sizable EDMs to be generated by
one-loop graphs when supersymmetric particle masses are below $\sim
1$ TeV.

However, the current experiment bounds on electron, neutron EDM, and
$^{199}$Hg atom are already  tight: $|d_e|<1.6\times 10^{-27} e
~{\rm cm}$ (90\% C.L.) \cite{Regan:ta}, $|d_n|< 2.9\times 10^{-27} e
~{\rm cm}$ (90\% C.L.) \cite{baker06}, and $|d_A(^{199}\textrm{Hg})|
< 2.1\times 10^{-27}e ~{\rm cm}$ (95\% C.L.) \cite{Romalis:2000mg}
(For recent reviews of EDM searches and their implications for SUSY,
see, {\em e.g.} Refs.~\cite{Pospelov:2005pr,RamseyMusolf:2006vr}).
These results imply CPV phases of order $10^{-3}$ or smaller,
leading to the so-called \lq\lq SUSY CP problem". Its resolution, as
well as that of the related \lq\lq SUSY flavor problem", requires
some other mechanism for suppressing one-loop EDMs (and one-loop
flavor changing neutral currents). One possibility is to take the
masses of the first and second generation sfermions to be of order
10 TeV \cite{Cohen:1996vb}. In such circumstances, the one-loop
contributions to EDMs are highly suppressed, and the two-loop
contributions to EDM, with CP violation from either
chargino-neutralino sector or the third generation of squarks, may
give competitive and even dominate contributions to EDMs of the
electron and neutron\footnote{The  EDMs of diamagnetic atoms such as
$^{199}$Hg will be suppressed in this limit, as they are generated
primarily by the one-loop chromo-EDM operators.}.

Previous work has considered a subset of these two-loop
contributions, including those involving third generation squarks
\cite{scalar,Chang:1999zw} and charginos
\cite{Chang:2002ex,Pilaftsis:2002fe,Giudice:2005rz,Chang:2005ac} whose CPV
interactions with the gauge-Higgs sector of the MSSM induce an EDM
(or chromo-EDM) of an elementary, first generation SM fermion. In particular, implications for CP violation at high-energy colliders and dominant higher-loop contributions were discussed in detail in Ref.~\cite{Pilaftsis:2002fe}. The CP
violation from chargino ($\chi^+$)-neutralino ($\chi^0$) sector can be propagated to the
SM fermion though purely gauge boson exchanges. In this case, it has
been shown that no CP violation can be propagated though $\gamma
\gamma$, $\gamma Z$, and $ZZ$ exchanges \cite{Giudice:2005rz},
leaving the $WW$ exchange as the only possibility. This contribution was recently calculated in
Ref. \cite{Giudice:2005rz,Chang:2005ac}. CP violation can also be propagated through the exchange of gauge and Higgs boson pairs, including $\gamma h^0$, $\gamma H^0$, $Zh^0$, $ZH^0$, $\gamma A^0$, $ZA^0$, and $WH^{\pm}$. Here, $h^0$
and $H^0$ denote the neutral, CP-even Higgs scalars of the MSSM,
with $h^0$ being the lightest, \lq\lq SM-like" scalar; $A^0$ is the
neutral CP-odd scalar; and $H^\pm$ denotes the charged scalars. The
contributions due to $\gamma h^0$, $\gamma H^0$, $Zh^0$, $ZH^0$, and
$\gamma A^0$ exchanges have been studied
\cite{Chang:2002ex,Pilaftsis:2002fe,Giudice:2005rz,Chang:2005ac}.

In what follows, we compute the remaining two-loop contributions
that survive in the limit of large sfermion masses: Barr-Zee
\cite{Barr:1990vd} type amplitudes wherein chargino-neutralino loops
communicate CPV to the fermion via the exchange of a $ZA^0$ or
$WH^\pm$ pair. We also compute the $\gamma h^0$, $\gamma H^0$, $Z
h^0$, $Z H^0$, and $\gamma A^0$ contributions, and compare our
results with the previous computations reported in
Refs.~\cite{Chang:2002ex,Pilaftsis:2002fe,Giudice:2005rz,Chang:2005ac}. We report
agreement with all previous results. We find that, in general, the new
contributions are comparable in magnitude to those previously
computed, or they are even dominant. The $ZA^0$ contribution is accidentally suppressed in the
case of the electron EDM by the $1-4{\rm sin}^2 \theta_W$ factor,
but it is important for the neutron EDM. Unlike the case of two-loop
diagrams with CP violation from squarks, where it has been noted
that the $\gamma h^0$ and $\gamma A^0$ contributions dominate
\cite{scalar,Chang:1999zw}, we find -- after completing a numerical
study of the analytic results -- that the $WH^{\pm}$ contribution is among the dominant ones for the electron EDM, and both the $Z$ plus $H^0,\ A^0$
and the $WH^{\pm}$ contributions are the dominant ones for the
neutron EDM, proving that the inclusion of these contributions is
indispensable. 

Apart from the implications for EDM phenomenology, our results also
have interesting consequences for the viability of supersymmetric
EWB. Indeed, part of our original motivation for computing the loops
containing the $A^0$ and $H^\pm$ is that the masses of these scalars
affects the dynamics of supersymmetric EWB. In particular, during a
first order electroweak phase transition that proceeds via bubble
nucleation, the rate at which the neutral Higgs vacuum expectation
values (vevs) change across the bubble walls depends on $m_{A^0}$
(which also sets the scale for $m_{H^\pm}$). Since the CP-violating
asymmetries needed for baryon number production are generated during
the phase transition by scattering from these vevs, knowledge of the
bubble wall profiles and their dependence on the other MSSM
parameters is essential for determining the viability of
supersymmetric EWB. In general, SUSY EWB is enhanced for relatively
light $m_{A^0}$ -- a region in which the corresponding $ZA^0$, $ZH^0$
$\gamma A^0$, $\gamma H^0$ and $WH^\pm$ EDM contributions are also enhanced. In
our numerical study of the two-loop EDMs, we investigate the
corresponding $m_{A^0}$-dependence with an eye to these implications
for EWB.

Our discussion of these points is organized in the remainder of the
paper as follows. In Section~\ref{sec:twoloop} we provide details of
our two-loop computation and the analytic expressions for the
results. Section~\ref{sec:numerics} gives our numerical analysis. We
summarize our results in Section~\ref{sec:conclusions}, while
additional technical details are provided in the Appendix. We note
that, during the course of completing our study, a parallel
computation of the two-loop EDMs in SUSY using an effective field theory approach also
appeared~\cite{Feng:2008cn}. We comment on the similarities and
differences we find with that analysis.

\section{Two Loop EDMs: Higgs-Gauge mediated $\boldmath{\chi^+}$-$\boldmath{\chi^0}$ contributions}
\label{sec:twoloop}

\begin{figure}[t]
\begin{center}
\mbox{
\begin{picture}(0,220)(130,0)

\put(10,120){\insertfig{3.7}{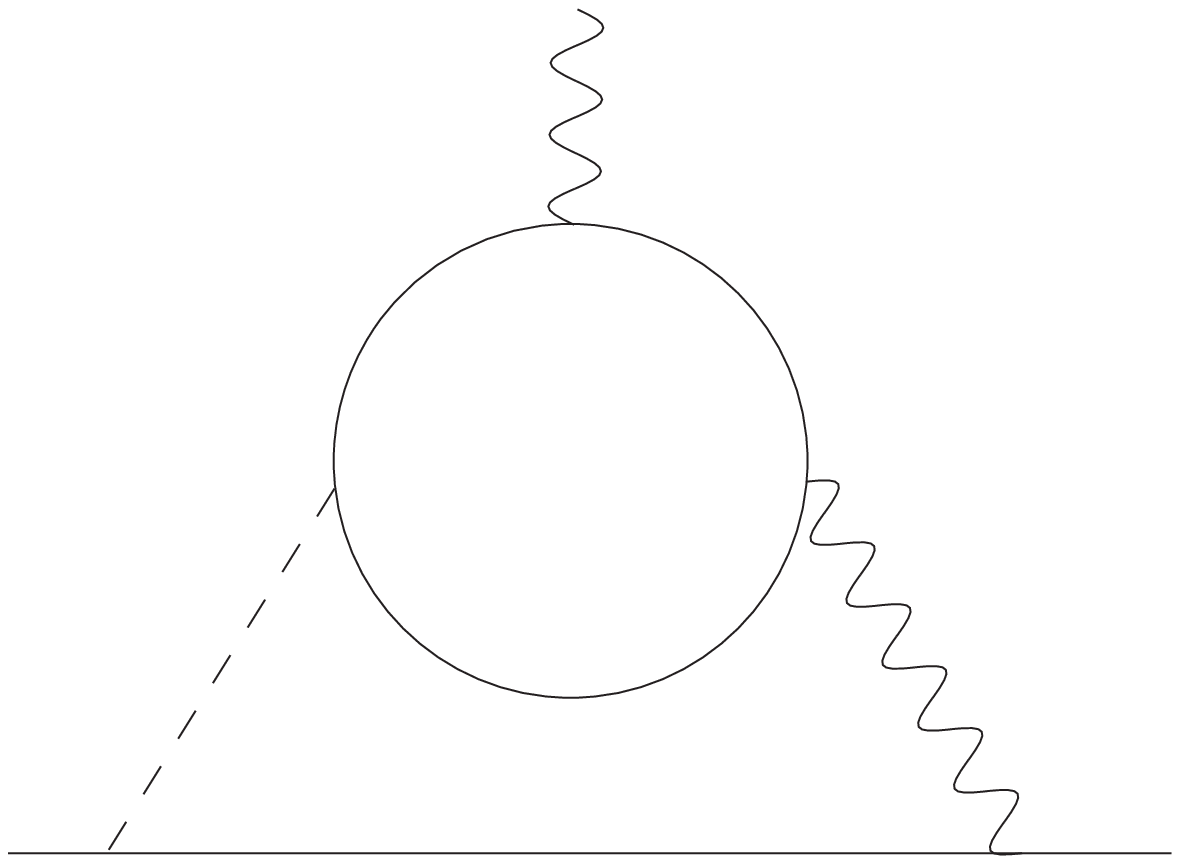}}
\put(150,120){\insertfig{3.7}{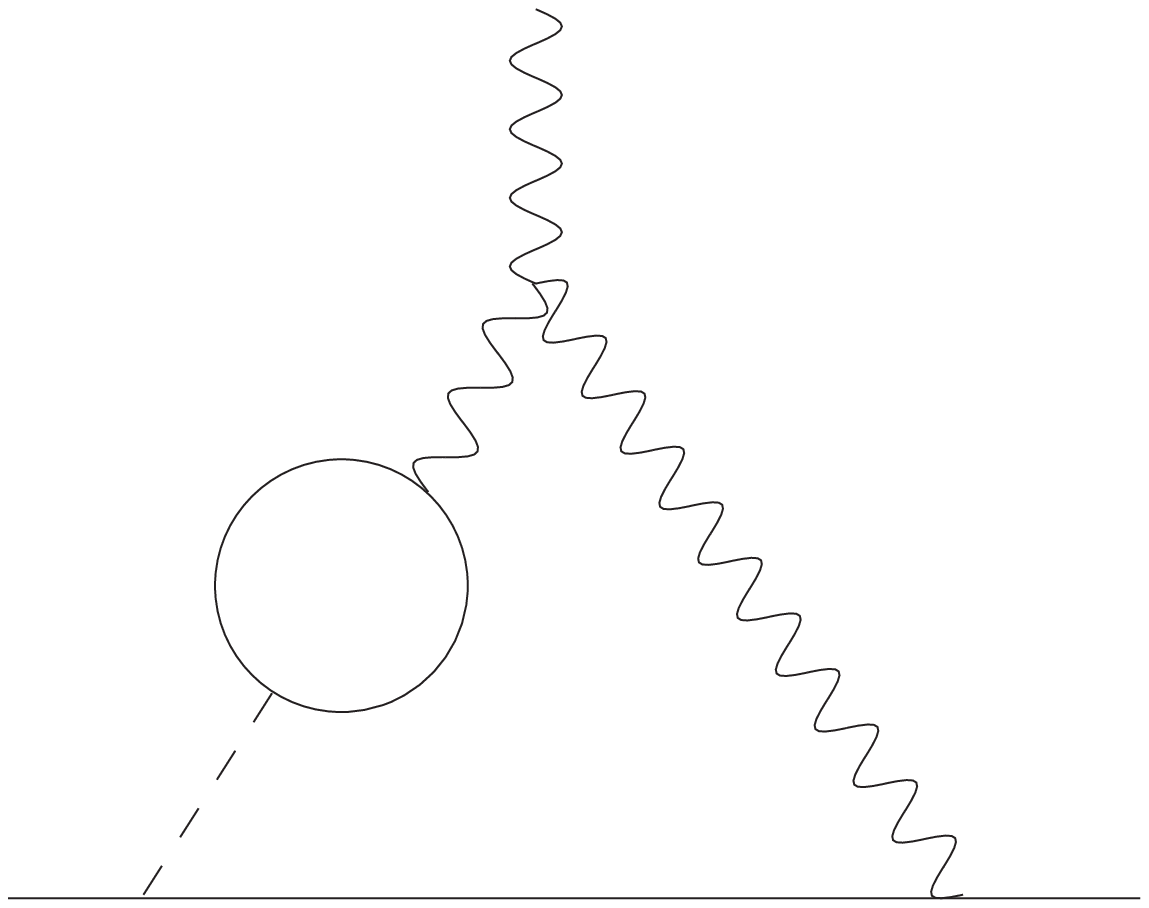}}
\put(10,10){\insertfig{3.7}{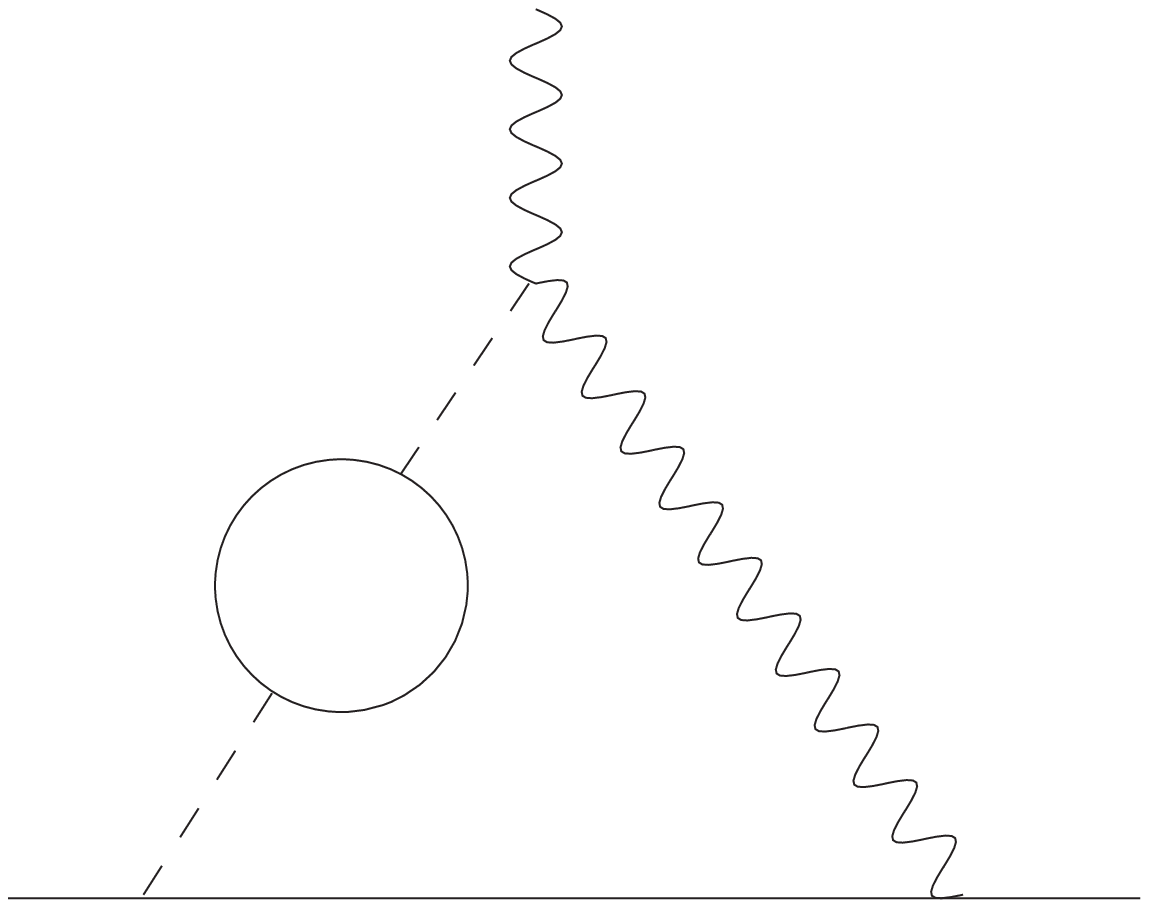}}
\put(150,10){\insertfig{3.7}{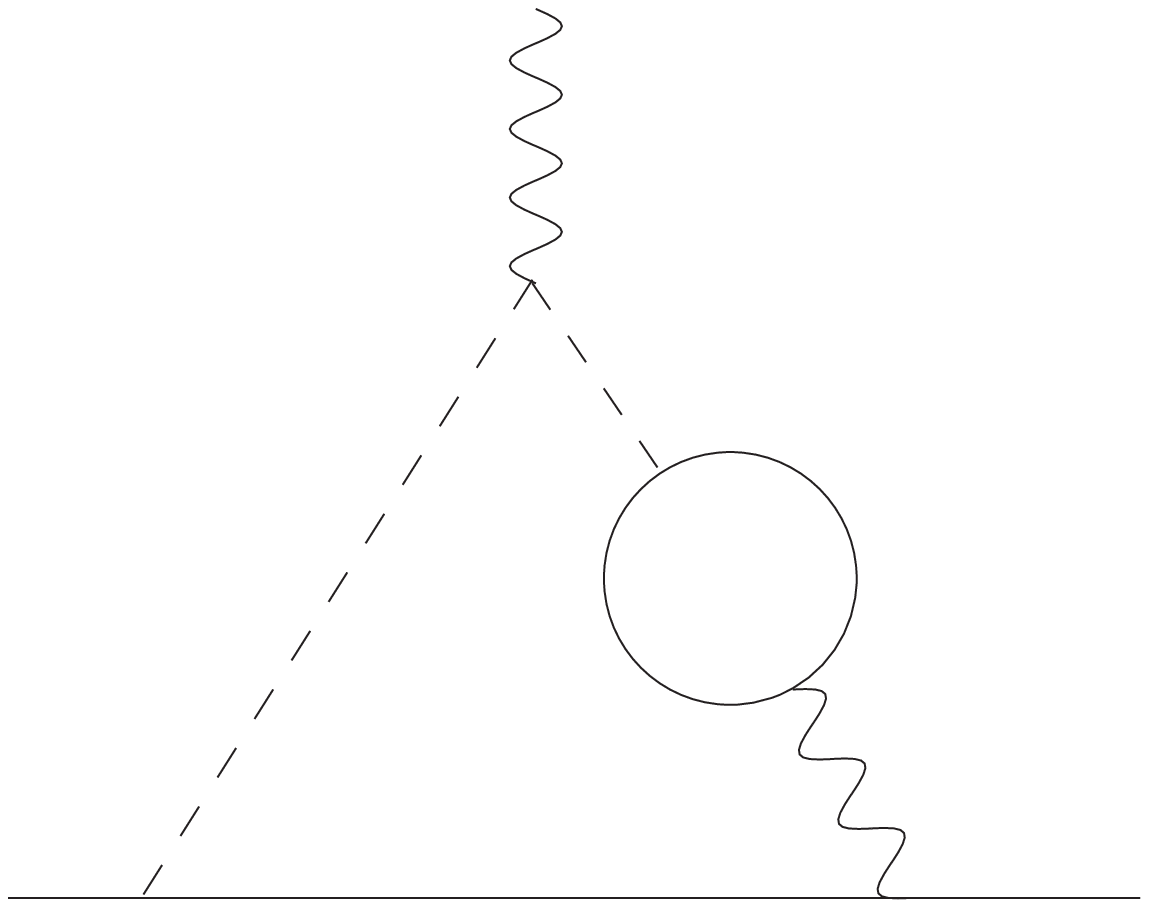}}

\Text(60,105)[c]{(a)} \Text(180,105)[c]{(b)} \Text(60,0)[c]{(c)}
\Text(180,0)[c]{(d)}

\Text(68,195)[c]{$\gamma$} \Text(208,200)[c]{$\gamma$}
\Text(68,90)[c]{$\gamma$} \Text(208,90)[c]{$\gamma$}

\Text(114,140)[c]{$\gamma$,Z,$W^+$} \Text(13,142)[c]{$h^0$,$H^0$,}
\Text(12,132)[c]{A,$H^+$} \Text(32,165)[c]{$\chi^+_a$}
\Text(90,165)[c]{$\chi^+_a$} \Text(62,128)[c]{$\chi^+_b,\chi^0_i$}

\Text(162,132)[c]{$H^+$} \Text(183,170)[c]{$W^+$}
\Text(235,150)[c]{$W^+$} \Text(164,160)[c]{$\chi^+_a$}
\Text(197,140)[c]{$\chi^0_i$}

\Text(20,25)[c]{$H^+$} \Text(24,49)[c]{$\chi^+_a$}
\Text(56,30)[c]{$\chi^0_i$} \Text(47,63)[c]{$G^+$}
\Text(93,43)[c]{$W^+$}

\Text(175,43)[c]{$H^+$} \Text(215,63)[c]{$H^+$}
\Text(199,31)[c]{$\chi^0_i$} \Text(235,53)[c]{$\chi^+_a$}
\Text(240,23)[c]{$W^+$}

\end{picture}
}
\end{center}
\caption{All the two-loop diagrams with chargino-neutralino loop
mediated by Higgs bosons. (Mirror graphs are not displayed.)
\label{fig:LoopGraphs}}
\end{figure}

Representative diagrams from the various topologies we consider are
shown in Fig. \ref{fig:LoopGraphs}. For simplicity we have not shown
the different crossed graphs or those in which scalar and vector
boson lines are interchanged. In addition, diagrams involving a
photon insertion on the SM fermion line are also not shown, as these
contributions vanish. Of the remaining, non-vanishing diagrams, the
contributions with neutral Higgs boson exchange, namely, $\gamma
h^0$, $\gamma H^0$, $Zh^0$, $ZH^0$, and $\gamma A^0$, $ZA^0$,  only
involve diagram (a), while the contributions with charged Higgs
boson exchange $W^{\mp}H^{\pm}$ involve all the diagrams (a,b,c,d).
To simplify the computation of the latter, we follow Ref.
\cite{Chang:1999zw} and use the nonlinear $R_{\xi}$ gauge
\cite{nonlinear}. The corresponding gauge-fixing term in the
Lagrangian is obtained by replacing the ordinary derivative that
appears in the  $R_{\xi}$ gauge
\begin{equation}
{\cal L}^{R_{\xi}}_{\textrm{g.f.}} = -\frac{1}{2\xi}|\partial_{\mu} W^+_{\mu}
- i \xi M_W \phi^+|^2,
\end{equation}
with the U(1)$_\textrm{EM}$ covariant derivative
$D_{\mu}=\partial_{\mu} - ie A_{\mu}$ in nonlinear $R_{\xi}$ gauge
\begin{equation}
{\cal L}^{\textrm{nonlinear}\, R_{\xi}}_{\textrm{g.f.}} = -\frac{1}{2\xi}|D_{\mu}
W^+_{\mu} - i \xi M_W \phi^+|^2\ \ \ .
\end{equation}
Thus, just as the $R_{\xi}$ gauge is designed to eliminate mixing
between the would-be Goldstone boson $G^{\pm}$ and the $W^\pm$
implied by the Higgs kinetic term $(D_{\mu}\phi)^+ (D^{\mu}\phi)$,
the nonlinear $R_{\xi}$ gauge is constructed in such a way that the
coupling $G^{\pm}W^{\mp}\gamma$ arising from the same kinetic term
is also canceled. As a result, the $W^+W^-\gamma$ vertex is modified
from its standard form in the conventional renormalizable gauges.

A direct -- and simplifying --  consequence of employing the nonlinear $R_{\xi}$ gauge is that the
contribution from diagram (c) vanishes due to the absence of the
$G^{\pm}W^{\mp}\gamma$ coupling. Moreover, an additional simplification can be obtained when carrying out the computation in the Landau gauge ($\xi\rightarrow0$). In doing so, one must take care to first compute the $\xi\not=0$ contributions to the $W^+W^-\gamma$ vertex and $W$-boson propagators in Fig. \ref{fig:LoopGraphs}(b) and  carry out the appropriate contractions that appear in the one-loop amplitude before taking the $\xi\to 0$ limit, since the additional contribution to the $W^+W^-\gamma$ vertex arising in the nonlinear $R_\xi$ gauge is proportional to $1/\xi$. The resulting simplification is that the amplitude from Fig. \ref{fig:LoopGraphs}(d) vanishes as well.  This is because the
chargino-neutralino loop in (d) is proportional to the four momentum
of $W$ boson, while the propagator of $W$ is transverse in Landau
gauge.

In carrying out the calculation, we first compute out the one-loop sub-graphs corresponding to the amplitude for
$\gamma(q,\mu) \rightarrow h(q-\ell) + g(\ell,\nu)$, where $h$ stands for
one of Higgs bosons ($h^0,H^0,A^0,H^{\pm}$) having momentum $q-\ell$;  $g$ denotes  one of
gauge bosons ($\gamma, Z, W$) having momentum $\ell$; and $\mu$ and $\nu$ denote the vector indices associated with the external photon and $g$, respectively.  Gauge invariance implies that the amplitude involves a linear combination of the (pseudo) tensors
\begin{eqnarray}
\label{eq:tensors}
P^{\mu\nu} & = & \epsilon^{\mu\nu\alpha\beta}q_{\alpha}\ell_{\beta}\\
\nonumber
T^{\mu\nu} & = & \ell^{\mu}q^{\nu}-g^{\mu\nu}\ell\cdot q\ \ \ .
\end{eqnarray}
For the full two-loop graphs involving the exchange of neutral
bosons, only $P^{\mu\nu}$ contributes in the case of CP-even Higgs
exchange, while only $T^{\mu\nu}$ contributes for the graphs
involving the CP-odd Higgs. Both $P^{\mu\nu}$ and $T^{\mu\nu}$
contribute to the two-loop $WH^{\pm}$ amplitude. For our particular
gauge choice, we find that $P^{\mu\nu}$ arises from diagram  Fig.
\ref{fig:LoopGraphs}(a) alone, while for $W^{\mp}H^{\pm}$ exchange, $T^{\mu\nu}$ requires the sum
of both Fig. \ref{fig:LoopGraphs}(a) and (b) [graph (b) only
generates a $\ell\cdot q g^{\mu\nu}$ structure].  We will use these
features to explain the origin of the overall, relative signs
between the various contributions below.

In obtaining our final results for the two-loop contributions, we
use the Feynman rules and conventions given in
Ref.~\cite{Rosiek:1995kg}\footnote{However, our convention for the Higgs scalar mixing angle
$\alpha$ differs from that of Ref.~\cite{Rosiek:1995kg}. To facilitate comparison with results appearing in the
literature, we adopt the convention of Ref.~\cite{Chang:2005ac}.}.
We have attempted to express our results
in a manner that makes easy to directly compare with the earlier work of
Refs.~\cite{Chang:2002ex,Pilaftsis:2002fe,Giudice:2005rz,Chang:2005ac}. We find
\begin{eqnarray}
\label{eq:gammaS}
d^{\gamma S}_f&=&\frac{e Q_f \alpha^2 c^{S}_f}{8 \sqrt{2} \pi^2
s^2_W } \frac{m_f }{M_W m^2_{S}} \sum^2_{a=1} {\rm Im}(D^R_{S,aa})
M_{\chi^+_a} \int^1_0
dx\frac{1}{x(1-x)}j(0,\frac{r_{aS}}{x(1-x)}),
\end{eqnarray}
\begin{eqnarray}
\label{eq:ZS}
d^{ZS}_f&=&\frac{e \alpha^2  (T_{3f_L}-2s^2_W Q_f)c^{S}_f}{16
\sqrt{2} \pi^2 c^2_W s^4_W } \frac{m_f }{M_W m^2_{S}}  \nonumber \\
&& ~~~~\times
 \sum^2_{a,b=1} {\rm Im}(G^R_{ab}D^R_{S,ba} -
 G^L_{ab}D^L_{S,ba}) M_{\chi^+_b}
 \int^1_0
dx\frac{1}{x}j(r_{ZS},\frac{xr_{aS}+(1-x)r_{bS}}{x(1-x)}),
\end{eqnarray}
\begin{eqnarray}\label{eq:gammaA}
d^{\gamma A^0}_f&=& \frac{e Q_f \alpha^2 c^{A^0}_f}{8\sqrt{2} \pi^2
s^2_W} \frac{m_f }{M_W m^2_{A^0}} \sum^2_{a=1} {\rm Im} E^R_{aa}
M_{\chi^+_a} \int^1_0 dx
\frac{1-2x+2x^2}{x(1-x)}j(0,\frac{r_{aA^0}}{x(1-x)}),
\end{eqnarray}
\begin{eqnarray}
\label{eq:ZA} d^{ZA^0}_{f}&=&\frac{e \alpha^2  (T_{3f_L}-2s^2_W
Q_f)c^{A^0}_f}{16 \sqrt{2} \pi^2 c^2_W s^4_W} \frac{m_f }{M_W
m^2_{A^0}} \nonumber \\
&& ~~~~ \times \sum^2_{a,b=1} {\rm Im}(G^R_{ab}E^R_{ba} +
G^L_{ab}E^L_{ba}) M_{\chi^+_b}
  \int^1_0
dx\frac{1-x}{x}j(r_{ZA^0},\frac{xr_{aA^0}+(1-x)r_{bA^0}}{x(1-x)}),
\end{eqnarray}
\begin{eqnarray}
\label{eq:WH} d^{WH^\pm}_{f} &=& -\frac{e \alpha^2  c^{H^+}_f}{32 \pi^2
s^4_W c_W } \frac{m_f}{M_W m^2_{H^+}} \sum^2_{a=1} \sum^4_{i=1}
\int^1_0 dx
\frac{1}{1-x}j(r_{WH^+},\frac{r_{aH^+}}{1-x} + \frac{r_{iH^+}}{x})  \nonumber \\
&&~~~ \left[ ({\rm Im}(M^L_{ai} N^{L*}_{ai} + M^R_{ai} N^{R*}_{ai})
M_{\chi^+_a} x^2  + {\rm Im}(M^R_{ai} N^{L*}_{ai} + M^L_{ai}
N^{R*}_{ai}) M_{\chi^0_i} (1-x)^2) \right.
 \nonumber \\
&& \left. + ({\rm Im}(M^L_{ai} N^{L*}_{ai} - M^R_{ai} N^{R*}_{ai})
M_{\chi^+_a} x
 + {\rm Im}(M^R_{ai} N^{L*}_{ai} - M^L_{ai}
N^{R*}_{ai}) M_{\chi^0_i} (1-x)) \right] \ \ \ .
\end{eqnarray}
Here,  $s_W={\rm sin}\theta_W$ and $c_W={\rm cos}\theta_W$. The $S$
in Eq. (\ref{eq:gammaS}) and (\ref{eq:ZS}) denotes $h^0$ and $H^0$. The symbol $f=u,d,e$ represents
the up quark, down quark, and electron, respectively; $Q_f$ and
$m_f$ are the electric charge and mass, respectively, of fermion
$f$, and $T_{3f_L}$ is the third component of the weak isospin of
its left-handed component; finally, $j(r,r')$ is the loop function
defined in Ref.~\cite{Giudice:2005rz} and given in the Appendix.

The mass ratios in loop functions are $r_{Zh^0}=M^2_Z/m^2_{h^0}$,
$r_{ZH^0}=M^2_Z/m^2_{H^0}$, $r_{ZA^0}=M^2_Z/m^2_{A^0}$,
$r_{ah^0}=M^2_{\chi^+_a}/m^2_{h^0}$,
$r_{aH^0}=M^2_{\chi^+_a}/m^2_{H^0}$,
$r_{aA^0}=M^2_{\chi^+_a}/m^2_{A^0}$, $r_{WH^+}=M^2_W/m^2_{H^+}$,
$r_{aH^+}=M^2_{\chi^+_a}/m^2_{H^+}$, and
$r_{iH^+}=M^2_{\chi^0_i}/m^2_{H^+}$, with $M_{W,Z}$, the masses of
$W$ and $Z$ gauge bosons, $m_{h^0,H^0,A^0,H^+}$, the masses of Higgs
bosons, and $M_{\chi^+_a}\geq0$ and $M_{\chi^0_i}\geq0$, the masses
of charginos and neutralinos, respectively. The coefficients
$c^{h^0,H^0,A^0,H^+}_{u,d,e}$ and matrices
$D^{R,L}_{h^0,H^0},G^{R,L},E^{R,L},M^{R,L},N^{R,L}$ involve various
combinations of the chargino and neutralino couplings to Higgs and
gauge bosons, and are collected explicitly in the Appendix.

Before proceeding with our numerical study, we make several comments
on the analytic results.
\begin{itemize}
\item[(i)] The dependence on
the CPV phases in the gauge-gaugino-Higgs-Higgsino sector is
contained in the imaginary parts of the couplings $D^R_{S,\, aa}$,
{\em etc.} but not separated out explicitly. As indicated in the
Appendix, these phases arise from diagonalizing the chargino and
neutralino mass matrices in Eq.~(\ref{eq:massmatrix}). In general,
the resulting independent phases are $\mathrm{Arg} (\mu M_i b^*)$ and
$\mathrm{Arg}(M_i M_j^*)$, where $\mu$ is the supersymmetric
Higgs-Higgsino mass parameter; $M_i$ ($i=1,2,3$) are the soft
gaugino mass parameters; and $b$ is the soft Higgs mass parameter.
The SU(3$)_C$ mass parameter does not enter the diagonalization of
chargino-neutralino mass matrices, leaving two remaining phases. The
analysis of CPV in this sector is often simplified by assuming that
$\mathrm{Arg}(M_1 M_2^*)=0$, leaving one remaining, independent
phase denoted $\phi_\mu$. In our numerical study below, we will
adopt this simplifying assumption and verify numerically that each
of the two-loop contributions is proportional to $\sin\phi_\mu$. We also comment on the
impact of relaxing this assumption.

\item[(ii)] The coefficients $c_f^{h^0}$ and $c_f^{H^0}$
given in Eq.~(\ref{eq:couplings}) as well as the matrices $D^{R,L}_{h^0}$ {\em
etc.} given in Eq.~(\ref{eq:matrices}) depend in general on
$\tan\beta=v_u/v_d$, where the $v_k$ are the vacuum expectation
values of the two neutral Higgs scalars, and $Z_R$, which further
depends on $m_{A^0}$ as illustrated in Eq.~(\ref{eq:dependence}).
This introduces an additional dependence on $\tan\beta$ and
$m_{A^0}$ beyond the explicit dependence generated by the Yukawa
couplings and dependence of the loop functions on scalar masses.

\item[(iii)] The overall sign in the expression for the $WH^\pm$ contribution is
opposite to that of the other contributions. The origin of this overall
sign can be understood by considering the combinations of couplings
and Lorentz structures entering the two loop amplitudes. To
illustrate, we consider the loop of Fig. \ref{fig:LoopGraphs}(a)
that enters each of the contributions. To compare the relative signs
of the couplings, we define a general set of interactions involving
charginos, neutralinos, Higgs scalars, and fermions:
\begin{eqnarray}
\nonumber
\mathcal{L}_{V\chi\chi} & = & {\bar\chi}\gamma^\mu \left[A_L P_L + A_R P_R\right]\chi V_\mu +\cdots\\
\nonumber
\mathcal{L}_{\phi\chi\chi} & = & {\bar\chi} \left[B_L P_L + B_R P_R\right]\chi \phi +\cdots\\
\label{eq:genint}
\mathcal{L}_{V\ell\ell} & = & {\bar\ell}\gamma^\mu \left[C_L P_L + C_R P_R\right]\ell V_\mu +\cdots\\
\nonumber
\mathcal{L}_{\phi\ell\ell} & = & {\bar\ell} \left[D_L P_L + D_R P_R\right]\ell \phi +\cdots
\end{eqnarray}
In Table \ref{tab:cases} below we give the corresponding phases of
the couplings obtained from the Feynman rules of
Ref.~\cite{Rosiek:1995kg}.
\begin{table}[!b]
\caption{Phases of couplings in Eq.~(\ref{eq:genint}) as they enter the amplitude of Fig. \ref{fig:LoopGraphs}(a) }
\begin{center}
\label{tab:cases}
\begin{tabular}{|c|c|c|c|}
\hline
Vertex & $ZA$ & $Zh^0$ & $WH^\pm$\\
\hline
$V\chi\chi$ & $-i$ & $-i$ & $+i$  \\
$\phi \chi\chi $ & $+1$ & $-i$ & $+i$ \\
$\phi\ell\ell$ & $+1$ & $+i$ & $-i$ \\
$V\ell\ell$ & $+i$ & $+i$ & $+i$ \\
\hline
Overall & $+1$ & $+1$ & $ -1$ \\
\hline
\end{tabular}
\end{center}
\end{table}
Now we consider the structure of the fermion line in the loop, which
gives the only other source of a phase difference between the
different contributions. If $\ell$ is the momentum flowing through
the loop (we may neglect the external fermion momenta for this discussion),
we have
\begin{eqnarray}
\nonumber
ZA^0 & \sim & \gamma^\mu\left[2T_3 P_L -2 Q\sin^2\theta_W \right]\not\!\ell \gamma_5 \\
\nonumber
&=& \gamma^\mu\left[- 2T_3 P_L +2 Q\sin^2\theta_W\gamma_5 \right]\not\!\ell \\
\nonumber
&=& \frac{1}{2}\gamma^\mu\left[-\left(2T_3  -4 Q\sin^2\theta_W\right)\gamma_5 -2 T_3\right]\not\!\ell
 \\
\label{eq:fermionline}
Zh^0 & \sim  & \gamma^\mu\left[2T_3 P_L -2 Q\sin^2\theta_W \right]\not\!\ell  \\
\nonumber
& = & \frac{1}{2}\gamma^\mu\left[\left(2T_3  -4 Q\sin^2\theta_W\right) -2 T_3 \gamma_5 \right]\not\!\ell\\
\nonumber
WH^\pm & \sim &  \gamma^\mu P_L \not\!\ell P_R = \gamma^\mu P_L \not\!\ell =\frac{1}{2} \gamma^\mu \left(1-\gamma_5\right) \not\!\ell
\end{eqnarray}
For  the case of $ZA^0$ exchange which involves $T^{\mu\nu}$ from
the closed chargino loop, we require the $\gamma_5$ term from the
lower line to obtain the EDM, whereas for $Zh^0$ exchange, we have
the pseudo-tensor $P^{\mu\nu}$ from the closed chargino loop,
necessitating that we retain the identity matrix term from the fermion line. For
the $WH^\pm$ exchange contribution, we require both. Table
\ref{tab:overall} gives the resulting overall phase for the various
contributions.

\begin{table}[!b]
\caption{Summary  of signs from fermion line and overall result. The final row is obtained by multiplying the overall phases from Table \ref{tab:cases} and the sign obtained from the fermion line. }
\begin{center}
\label{tab:overall}
\begin{tabular}{|c|c|c|c|c|}
\hline
Graph & $ZA$ ($T^{\mu\nu}$) & $Zh^0$ ($P^{\mu\nu}$)& $WH^\pm$ ($T^{\mu\nu}$) & $WH^\pm$ ($P^{\mu\nu}$)\\
\hline
$\gamma$-matrix & $\gamma_5$ & $1$ & $\gamma_5$ & $1$  \\
sign  & $-1$ & $+1$ & $-1$ & $+1$ \\
\hline
Overall & $-1$ & $+1$ & $+1$ & $-1$ \\
\hline
\end{tabular}
\end{center}
\end{table}
We observe that the expression for the part of the $WH^\pm$ exchange graph arising from the $T^{\mu\nu}$ tensor should have an opposite, overall phase compared to the corresponding term for the $ZA$ exchange diagram. Similarly, the $P^{\mu\nu}$ component of the $WH^\pm$ loop and the $Zh^0$ graph will also differ in overall relative phase. Note that there is an additional overall phase that arises between the $T^{\mu\nu}$ and $P^{\mu\nu}$ terms when the identity
\begin{equation}
\nonumber
\varepsilon^{\mu\nu\alpha\beta} \sigma_{\alpha\beta} = -2i \sigma^{\mu\nu}\gamma_5
\end{equation}
is used in the terms generated by $P^{\mu\nu}$. Thus, the relative
sign between the $ZA^0$ and $Zh^0$ graphs are the same, as are the
relative phase between the $T^{\mu\nu}$ and $P^{\mu\nu}$ terms in
the $WH^\pm$ contribution. As we discuss below, the resulting overall sign in Eq.~(\ref{eq:WH}) is compensated by the signs of various matrix element combinations $M_{ai}^L N_{ai}^{L\ast}$ {\em etc.} that enter the sum over chargino-neutralino intermediate states in the $\tan\beta\gtrsim 1$ regime. Consequently, the $WH^\pm$ and other gauge boson-scalar exchange contributions to the electron and down-quark EDMs add coherently in the phenomenologically allowed regions of MSSM parameter space. 

\item[(iv)]Our results for the $\gamma h^0$,  $\gamma
A^0$, and $Zh^0$ amplitudes agree with those of
Ref.~\cite{Chang:2002ex,Pilaftsis:2002fe,Giudice:2005rz}, including the overall
phase. On the surface, our result for the $Zh^0$ contribution
appears to be different  from the expression given in Ref.~\cite{Giudice:2005rz}. The difference amounts to
replacing the $x^{-1}$ in Eq.~(\ref{eq:ZS}) by $[2x(1-x)]^{-1}$ to
convert the integral to that of Ref.~\cite{Giudice:2005rz}. However,
after taking into account the symmetry properties of the integrands in both expressions, 
we have verified (both analytically and numerically) that they agree.

\item[(v)] A direct comparison of our analytic results with those obtained in
Ref.~\cite{Feng:2008cn} is not straightforward, since the latter employed an effective field theory
approach. We note, however, that these authors also include a nonzero result for the $ZZ$-exchange
contribution that the authors of Ref.~\cite{Giudice:2005rz} argued should vanish. A direct comparison of numerical
results is also challenging, since only the dependence of the EDMs on the CPV phases was given in Ref.~\cite{Feng:2008cn}, whereas in our numerical study below, we explore the dependence on mass parameters and $\tan\beta$ for fixed values of $\phi_\mu$.

\end{itemize}

\section{numerical analysis}
\label{sec:numerics}

In this section we numerically assess the impact of the additional,
two-loop EDM contributions discussed above. As mentioned in the
Introduction, one  motivation for our work to consider the complete
set of two-loop Higgs-mediated chargino-neutralino contributions
stems from the framework of EWB. It is therefore natural and
well-motivated to focus a portion of our numerical analysis on a supersymmetric
setup which is compatible with that framework. Before doing so,
however, we investigate the relative importance of the various
contributions and their dependence on MSSM parameters.

To that end, we define a benchmark parameter set scenario that will serve as a basis for comparison, motivated by the EWB framework, and consistent with phenomenological and cosmological constraints, as discussed below. We then proceed with a scan over pairs of parameters that govern the size of EDMs, keeping the other parameters fixed at their benchmark values. To suppress one-loop EDM contributions, we assume all sfermions to be decoupled (we set all sfermion soft breaking masses to 10 TeV, and the trilinear scalar couplings to zero, for definiteness). The gluino mass is entirely unimportant for the phenomenology discussed here, and is set to 1 TeV. The remaining  parameters relevant  for the two-loop EDM contributions are the absolute values of the gaugino soft breaking masses $M_{1,2}$ and of the higgsino mass parameter $\mu$, the heavy MSSM Higgs mass scale (for definiteness we employ here as a free parameter $m_{A^0}$), and $\tan\beta$. Our reference benchmark setup is defined as follows:
\begin{equation}
M_1=145 \ {\rm GeV},\ \ M_2=290 \ {\rm GeV},\ \  \mu=300 \ {\rm GeV},\ \  m_{A^0}=300 \ {\rm GeV},\ \  \tan\beta=10.\label{eq:refval}
\end{equation}

We consider here one single CP-violating phase, $\phi_\mu$, as discussed above. We set this phase $\phi_\mu=\pi/2$, giving the largest CP-violating effect. We verified numerically  that the EDMs considered here scale proportionally to $\sin\phi_\mu$ to within an accuracy of 1\% . This means that (1) all the results we show below can be simply re-scaled when assuming a non-maximal CP-violating phase and (2) we show the largest possible size for the EDM contributions we consider here\footnote{We remind the reader that EWB implies $\sin\phi_\mu\gtrsim10^{-2}$ \cite{Cirigliano:2006dg}.}. Notice that the values for $m_{h^0}$ and the mixing angle $\alpha$ (at the two loop level) as well as all the supersymmetric masses and mixing are obtained numerically through the FeynHiggs package \cite{Heinemeyer:1998yj}.

We have chosen this particular benchmark setup for several reasons. First, this choice is potentially compatible with successful EWB. Second, the lightest neutralino relic abundance is close to the observed cold dark matter density \cite{wmap}. If the relic neutralino abundance were larger than the cold dark matter density, a mechanism would be needed to dilute the relic abundance, with implications for EWB as well \cite{usinprep}. Third, the parameter values given above are consistent with collider searches, with precision electroweak data, including the muon anomalous magnetic moment, and with the inclusive branching ratio $b\to s\gamma$ \cite{Yao:2006px}. The latter constraint is particularly critical at low $m_{A^0}$, since contributions to $b\to s\gamma$ from the top-quark-$H^\pm$ loop can be sizable.

\begin{figure}
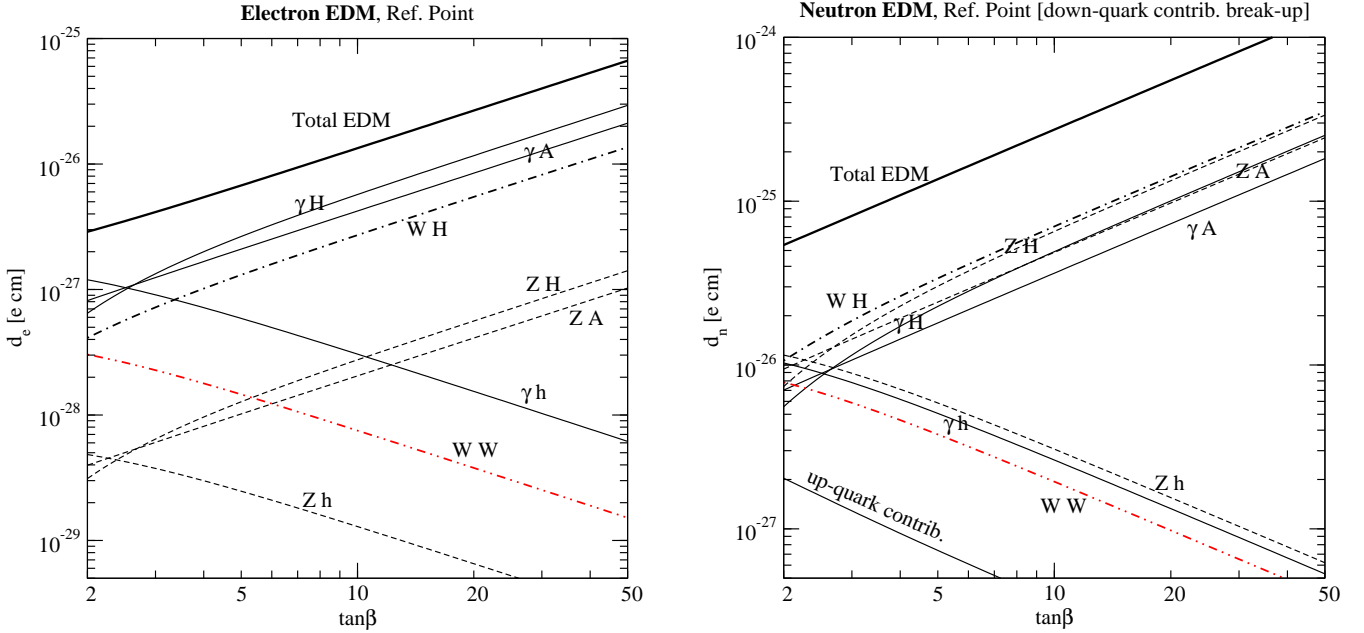

\mbox{\hspace*{-0.5cm}\includegraphics[width=8.5cm]{de_tanbe.eps}\qquad\includegraphics[width=8.5cm]{dn_tanbe.eps}}
\caption{A break-up of the various Higgs-mediated chargino-neutralino two-loop contributions to the electron (left) and neutron (right) electric dipole moment. Black lines correspond to positive values, red lines to negative values. In the right panel we show the various down-quark contributions times a factor 4/3, as well as the global up-quark contribution (times a factor -1/3). All SUSY parameters, except $\tan\beta$, are fixed to the reference setup.\label{fig:tanbe}}
\end{figure}

Starting from this reference point, we first illustrate the relative
magnitude of two-loop contributions in Fig.~\ref{fig:tanbe}. Here,
we explore these contributions to the EDM of  the
electron (left panel) and the neutron (right panel) as a function of $\tan\beta$, with
all other parameters set as in Eq.~(\ref{eq:refval}) . To
illustrate out results on a logarithmic scale, we show absolute
values and indicate with black lines positive contributions and with
red lines negative ones (the only negative contribution we find is from the $WW$ loop, for which we take the expression given in Ref. \cite{Giudice:2005rz}. 

For the computation of the neutron EDM from the quark EDM, we employ the na\"ive constituent quark model (CQM) relation
\begin{equation}
\label{eq:cqm}
d_n=\frac{4}{3}d_d-\frac{1}{3}d_u.
\end{equation}
The computation of $d_n$ is subject to considerable theoretical uncertainty associated with the non-perturbative strong interaction (for a discussion see, {\em e.g.}, Refs.~\cite{Pospelov:2005pr,Ellis:2008zy}. QCD contributions to the renormalization group evolution of the quark EDM operators from the weak scale to the hadronic scale lead to an overall enhancement factor of 1.53 that multiplies the combination of up- and down-quark EDMs on the right side or Eq.~(\ref{eq:cqm}) \cite{Ibrahim:1997gj}. Alternately, the use of QCD sum rule techniques to compute neutron matrix elements of the quark EDM operators leads to a multiplicative enhancement factor of $1.4\pm 0.6$ associated with the $q{\bar q}$ condensate\cite{Pospelov:2005pr}. Ellis and Flores have computed this matrix element by relating the quark EDM contributions to the corresponding quark contributions to the nucleon spin as implied by polarized deep inelastic scattering measurements, neutron $\beta$-decay, and the Bjorken sum rule\cite{Ellis:1996dg}. This approach leads to a different weighting of the up- and down-quark contributions than appears in the Eq.~(\ref{eq:cqm}) as well as a substantial strange quark contribution that is absent from the CQM and QCD sum rule computations. In  light of these variations, we consider Eq.~(\ref{eq:cqm}) to provide a reasonable benchmark, bearing in mind that a first principles (lattice) QCD computation may yield a different dependence on the light quark EDMs.

The resulting curves in Fig.~\ref{fig:tanbe} lead to several observations.
\begin{itemize}
\item[(i)] As indicated earlier, the sum over all intermediate chargino-neutralino states in Eq.~(\ref{eq:WH}) compensates for the overall relative sign in the expression for the $WH^\pm$ contribution. After analyzing the individual contributions in detail, we find that although the contribution from the lightest $\chi^+$-$\chi^0$ pair is negative (corresponding to the explicit sign in front of the expression), the sum is dominated by the remaining sets of intermediate states, many of which carry an opposite relative sign due to the chargino-neutralino mixing matrix elements. As a result, the $WH^\pm$ and other gauge boson-scalar exchange contributions to the electron and down-quark EDMs carry the same relative sign. 
\item[(ii)] All contributions involving heavy Higgses scale linearly with $\tan\beta$, while the $WW$ and the contributions involving $h^0$ scale as $1/\tan\beta$.
\item[(iii)] The dominant contributions to the electron EDM appear to be the $\gamma H^0$, $\gamma A^0$ and $WH^\pm$ loops. However notice that the $\gamma h^0$ contribution dominates at small $\tan\beta\sim2$, and that the $WW$ contribution is also sizable in that regime. The $Z$ plus Higgs contributions are suppressed by the $T_{3e_L}-2s_W^2Q_e$ factor, and are relatively subdominant. 
\item[(iv)] A similar picture applies to the case of the neutron EDM. Here we explicitly show only the break-up for the down-quark EDM contribution (times a factor 4/3), and the overall up-quark contribution (times a factor -1/3). As opposed to the electron EDM, the $ZH^0$, $ZA^0$, and $WH^\pm$ contributions dominate the down-quark EDM.
\end{itemize}

\begin{figure}
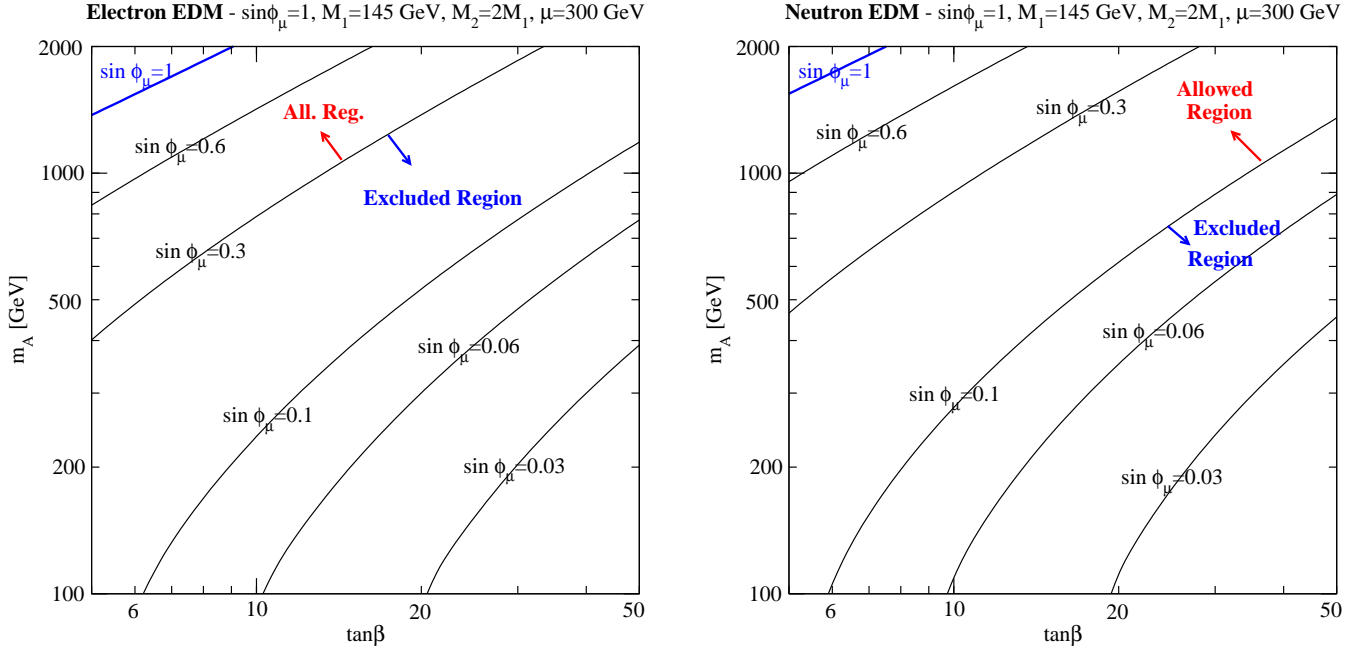

\mbox{\hspace*{-0.5cm}\includegraphics[width=8.5cm]{de_tb_ma_sphi.eps}\qquad\includegraphics[width=8.5cm]{dn_tb_ma_sphi.eps}}
\caption{Exclusion limits for the electron (left) and neutron (right) electric dipole moment, on the ($\tan\beta,m_{A^0}$) plane, for $M_1=145$ GeV, $M_2=290$ GeV, $\mu=300$ GeV and for various values of the CP violating phase $|\sin\phi_\mu|$. For each value of $|\sin\phi_\mu|$, parameter space points below the corresponding line are excluded, while those above the line are allowed.\label{fig:tbma}}
\end{figure}

Using the foregoing considerations, we now study the dependence of the total two-loop EDMs as a function of various MSSM parameters. First, we investigate the $(\tan\beta,m_{A^0})$ sector. This sets the mass scale for all loops involving $H^0$, $A^0$ and $H^\pm$, as well as various couplings, directly [see e.g. Eq.~(\ref{eq:couplings})] or indirectly, {\em e.g.}, through electroweak symmetry breaking effects in the neutralino and chargino mass and mixing matrices. We explore the EDM dependence on $(\tan\beta,m_{A^0})$ in Fig.~\ref{fig:tbma}, setting again all other supersymmetric parameters to the values indicated in (\ref{eq:refval}). We indicate in the figure the values of the CP violating phase $\sin\phi_\mu$ such that the resulting EDM equals the current experimental limit, $d_e=1.6\times 10^{-27}$ e cm (left) and $2.9\times 10^{-26}$ e cm (right). For each value of $|\sin\phi_\mu|$, parameter space points below the corresponding line are excluded, while those above the line are allowed, as indicated. As expected, we find a suppression of the EDM with increasing $m_{A^0}$ at fixed values of $\tan\beta$, thereby allowing for consistency between experimental limits and larger values of $|\sin\phi_\mu|$ . The overall behavior of the electron and the neutron EDM is remarkably similar. Presumably, this similarity  indicates that (1) the up-quark contribution to $d_n$ is subdominant (see Fig.~\ref{fig:tanbe}, right panel; we actually find that it features significant cancellations among the various contributions), and that (2) there are a few dominant contributions to $d_e$ and to $d_d$ that are simply proportional to each other, and hence they contribute in a similar fashion.

To make the connection with EWB, we now analyze the ($\mu$, $M_{1,2}$)-dependence of the two-loop EDMs. A generic expectation of the EWB scenario for the MSSM particle spectrum includes a relatively light mass scale for the heavy MSSM Higgs sector. This scenario depends on the suppression of the net baryon number density generated at the EW phase transition with $m_{A^0}$, as pointed out and quantified {\em e.g.} in Ref.~\cite{Moreno:1998bq}. It was also realized in several analyses (see {\em e.g.} Ref.~\cite{Carena:2002ss,Cirigliano:2006dg,usinprep} and references therein), that the requirement of sufficiently large CP-violating sources for successful EWB prefers a resonant enhancement in the higgsino-gaugino sources. This resonance occurs for   $M_{1}\sim\mu$, the resonant neutralino baryogenesis funnel, or $M_2\sim\mu$, the resonant chargino baryogenesis funnel, with  $M_{1,2},\mu\lesssim1$ TeV. An additional resonance could occur for the CPV  stop sources, but the latter possibility is generally precluded by the LEP limits on the mass of the $h^0$.  In both cases, however, successful EWB implies sub-TeV masses for higgsinos and gauginos.

Another generic feature of the MSSM spectrum implied by successful EWB is a light, mostly right-handed stop, in order to make the EW phase transition more strongly first order. Several studies pointed out, however,  that an extended, non-minimal Higgs sector can also significantly (and perhaps more naturally) enhance the first-order character of the EW phase transition \cite{singlet,Profumo:2007wc}. We thus do not regard the requirement of a light stop as a necessary feature of a supersymmetric setup giving successful EWB. From the viewpoint of EDMs, we considered and evaluated the size of the two-loop stop-mediated contributions in \cite{Cirigliano:2006dg}, and concluded that they are subdominant, provided the left-handed stop is heavy enough, as required for EWB in the MSSM\footnote{Note that this requirement stems primarily from having a Higgs mass consistent with the LEP limits}. In the interest of singling out the two-loop Higgs-mediated chargino-neutralino contributions under investigation here, and in view of the above considerations, we do not  assume a light right handed stop.

\begin{figure}
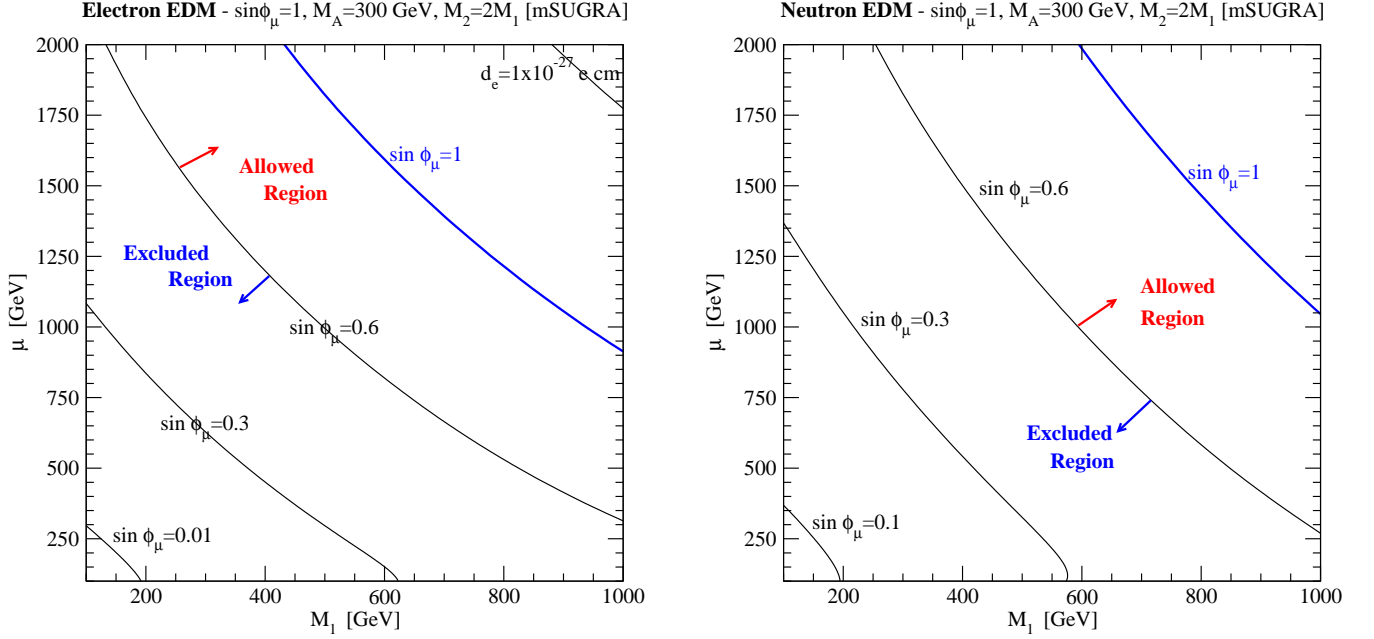

\mbox{\hspace*{-0.5cm}\includegraphics[width=8.5cm]{de_m1_mu_sphi.eps}\qquad\includegraphics[width=8.5cm]{dn_m1_mu_sphi.eps}}
\caption{Exclusion limits for the electron (left) and neutron (right) electric dipole moment, on the ($M_1,\mu$) plane. We assume the gaugino unification mass relation $M_2\simeq2M_1$, and vary the CPV phase $\sin\phi_\mu$ (see the text for further details on the model assumptions). In the upper right corner of the left panel we also show the contour of electron EDM equal to $10^{-27}$ e cm, for maximal CPV phase.\label{fig:m1mu}}
\end{figure}

\begin{figure}
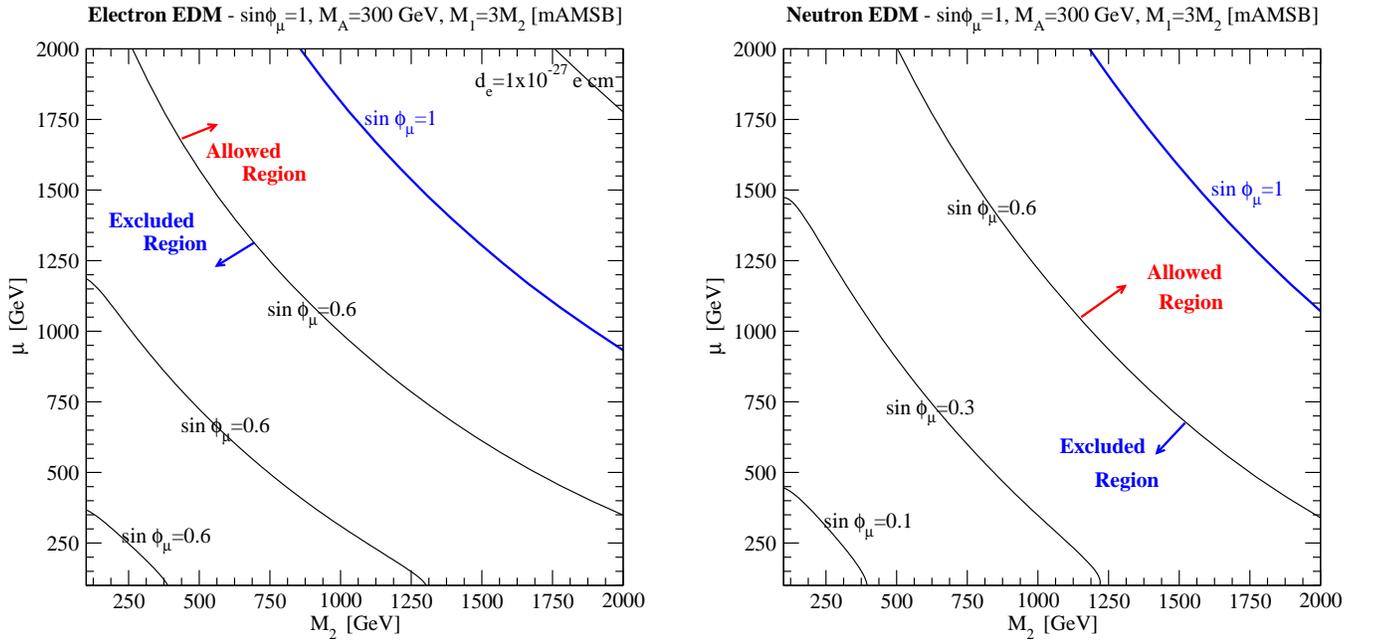

\mbox{\hspace*{-0.5cm}\includegraphics[width=8.5cm]{de_m2_mu_sphi.eps}\qquad\includegraphics[width=8.5cm]{dn_m2_mu_sphi.eps}}
\caption{Exclusion limits for the electron (left) and neutron (right) electric dipole moment, on the ($M_2,\mu$) plane. We assume here the anomaly-mediated SUSY breaking gaugino mass relation $M_1\simeq3M_2$. In the upper right corner of the left panel we also show the contour of electron EDM equal to $10^{-27}$ e cm, for maximal CPV phase.\label{fig:m2mu}}
\end{figure}

Having these considerations in mind, we illustrate in
Fig.~\ref{fig:m1mu}  the values of the CPV phase giving rise to an electron (left) and neutron (right) EDM equal to the current experimental limit, in the $(M_1,\mu)$ plane. We assume a minimal supergravity-type relation\footnote{Notice that the label ``[mSUGRA]'' obviously doesn't refer to the usual minimal supergravity setup, but only to the gaugino mass relation being employed here. The same applies to the label ``[mAMSB]'' of fig.~\ref{fig:m2mu}.} between the gaugino soft breaking masses (where gaugino masses unify at the GUT scale, and their EW-scale values are set by renormalization group running), and set here $M_2=2M_1$. In addition, $\tan\beta$ and $m_{A^0}$ are set to the reference values listed in Eq.~(\ref{eq:refval}). In the low $\mu$ and low $M_1$ region the size of the two-loop contribution exceeds the experimentally viable values (which we take to be $d_e<1.6\times 10^{-27}$ e cm and $d_e<2.9\times 10^{-26}$ e cm)  for a maximal CPV phase. For each value of the CPV phase, we indicate the boundaries of the excluded region. The parameter space above the various lines is currently experimentally open even for $\sin\phi_\mu=1$.

In passing, we note that a misalignment of the relative CPV phase between $M_1$ and $M_2$ does not affect our numerical results. In particular, we find that the main driver for the 2-loop EDM we consider here is the relative phase between $M_2$ and $\mu$, while a negligible contribution originates from the relative phase between $M_1$ and $\mu$. This result can potentially have profound implications for the interplay between electro-weak baryogenesis and EDM searches, which we plan to explore in a future study.

In summary, our numerical results indicate that the new contributions computed here are dominant for the electron and neutron EDM at the two-loop level (in the limit of heavy sfermions) and should thus be included in any study of EDMs and CP violation in the MSSM.

\section{summary and conclusions}
\label{sec:conclusions}

The analysis we have completed gives the EDMs of the electron and neutron in the MSSM when the sfermion masses are large, leading to a suppression of the one-loop contributions and general dominance of two-loop terms. Considering this regime allows one to circumvent the SUSY CP problem associated with the present, stringent EDM limits and sub-TeV scale sfermions. It also implies vanishing EDMs for diatomic atoms, assuming they are generated primarily by the chromo-EDMs of the quarks\footnote{See Ref.~\cite{Pospelov:2005pr} for a detailed discussion of the effective operator-dependence of various EDMs.}.  Previous studies of this regime have generally also taken all but the lightest, SM-like CP-even scalar to be heavy, thereby suppressing  two-loop contributions involving the other Higgs scalars as well. In the present study, we have not made these assumptions and have, instead, analyzed the dependence of the two-loop EDMs on the full gauge-gaugino-Higgs-Higgsino parameter space of the MSSM.

Our primary result is that contributions arising from
exchanges involving one SM gauge boson and either the CP-odd neutral
scalar, $A^0$,  or the charged Higgs scalars, $H^\pm$, are comparable or dominant to previously considered contributions. We have also
analyzed the prospective implications for MSSM electroweak
baryogenesis, whose viability depends in part on the values of
$\tan\beta$ and $m_{A^0}$. We leave a thorough exploration of the MSSM EWB to a more comprehensive future study \cite{usinprep}.

\section*{Acknowledgements}
We thank G. Giuidice and A. Romanino for helpful discussions regarding the results for the $Zh^0$ contributions. 
This work was supported in part under  U.S. Department of Energy Contracts DE-FG02-08ER41531(YL and MJRM)  and the University of Wisconsin Alumni Research Foundation (YL and  MJRM), and by a Faculty Research Grant from the University of California, Santa Cruz (SP).

\appendix
\label{appendix}

\section{coefficients, matrices, and loop function}

The coefficients $c^{h^0,H^0,A^0,H^+}_{u,d,e}$ depend on specific
types of Higgs bosons and SM fermions
\begin{eqnarray}
&&c^{h^0}_u=\frac{Z^{21}_R}{{\rm
sin}\beta},~~c^{h^0}_d=c^{h^0}_e=\frac{Z^{11}_R}{{\rm
cos}\beta},~~c^{H^0}_u=\frac{Z^{22}_R}{{\rm
sin}\beta},~~c^{H^0}_d=c^{H^0}_e=\frac{Z^{12}_R}{{\rm
cos}\beta},~~\nonumber \\
&&c^{A^0}_u={\rm cot}\beta,~~c^{A^0}_d=c^{A^0}_e={\rm
tan}\beta,~~c^{H^+}_u={\rm cot}\beta,~~c^{H^+}_d=c^{H^+}_e={\rm
tan}\beta.\label{eq:couplings}
\end{eqnarray}

The matrices $D^{R,L}_{h^0,H^0},G^{R,L},E^{R,L},M^{R,L},N^{R,L}$ are
\begin{eqnarray}
D^R_{h^0,ab} &=& Z^{11}_R Z^{2b*}_-Z^{1a*}_+ +
Z^{21}_R Z^{1b*}_-Z^{2a*}_+, \nonumber \\
D^L_{h^0,ab} &=& Z^{11}_R Z^{2a}_-Z^{1b}_+ + Z^{21}_R
Z^{1a}_-Z^{2b}_+ ,
\nonumber \\
D^R_{H^0,ab} &=& Z^{12}_R Z^{2b*}_-Z^{1a*}_+ +
Z^{22}_R Z^{1b*}_-Z^{2a*}_+, \nonumber \\
D^L_{H^0,ab} &=& Z^{12}_R Z^{2a}_-Z^{1b}_+ + Z^{22}_R
Z^{1a}_-Z^{2b}_+ ,
\nonumber \\
G^R_{ab} &=& \frac{1}{2}(Z^{1a}_-Z^{1b*}_- + \delta^{ab}(c^2_W -
s^2_W)), \nonumber \\
G^L_{ab} &=& \frac{1}{2}(Z^{1a*}_+Z^{1b}_+ +
\delta^{ab}(c^2_W - s^2_W)), \nonumber \\
E^R_{ab} &=& {\rm sin}\beta Z^{2b*}_-Z^{1a*}_+ +
{\rm cos}\beta Z^{1b*}_-Z^{2a*}_+, \nonumber \\
E^L_{ab} &=& -({\rm sin}\beta Z^{2a}_-Z^{1b}_+ + {\rm cos}\beta
Z^{1a}_-Z^{2b}_+), \nonumber \\
M^R_{ai} &=& Z^{1a}_- Z^{2i*}_N + \frac{1}{\sqrt{2}} Z^{2a}_-
Z^{3i*}_N, \nonumber \\
M^L_{ai} &=& Z^{1a*}_+ Z^{2i}_N - \frac{1}{\sqrt{2}} Z^{2a*}_+
Z^{4i}_N, \nonumber \\
N^R_{ai} &=& -{\rm cos}\beta (\frac{1}{\sqrt{2}} Z^{2a*}_+
(Z^{1i*}_N s_W + Z^{2i*}_N c_W) + Z^{1a*}_+ Z^{4i*}_N c_W),
\nonumber \\
N^L_{ai} &=& {\rm sin}\beta (\frac{1}{\sqrt{2}} Z^{2a}_-(Z^{1i}_N
s_W + Z^{2i}_N c_W) - Z^{1a}_- Z^{3i}_N c_W), \label{eq:matrices}
\end{eqnarray}
with ${\rm tan}\beta=v_u/v_d$.

The $Z_{\pm,N}$ are diagonalization matrices of chargino and
neutralino mass matrices $Z^T_- M_C Z_+ = {\rm
Diag}(M_{\chi^+_1},M_{\chi^+_2})$, $Z^T_N M_N Z_N = {\rm
Diag}(M_{\chi^0_1},M_{\chi^0_2},M_{\chi^0_3},M_{\chi^0_4})$, with
$M_{\chi^+_a}>0$, $M_{\chi^0_i}>0$, and
\begin{eqnarray}
M_C&=&\begin{pmatrix} M_2 & \sqrt{2}M_W{\rm sin}\beta \\
\sqrt{2}M_W{\rm cos}\beta & \mu
\end{pmatrix}, \nonumber \\
\label{eq:massmatrix}
M_N&=&\begin{pmatrix} M_1 & 0 & -M_Z s_W {\rm cos}\beta &
M_Z s_W {\rm sin}\beta \\
0 & M_2 & M_Z c_W {\rm cos}\beta &
-M_Z c_W {\rm sin}\beta \\
-M_Z s_W {\rm cos}\beta & M_Z c_W {\rm
cos}\beta & 0 & -\mu \\
M_Z s_W {\rm sin}\beta & -M_Z c_W {\rm sin}\beta & -\mu & 0
\end{pmatrix}.
\end{eqnarray}

The $Z_R$ in $c^{h^0,H^0}$ and $D^{R,L}_{h^0,H^0}$ is the matrix
that diagonalize the mass matrix of CP-even neutral Higgs bosons
\begin{equation}
\sqrt{2}\begin{pmatrix} {\rm Re}[H^0_d] - v_d \\
{\rm Re}[H^0_u] - v_u
\end{pmatrix} = Z_R \begin{pmatrix} h^0 \\
H^0
\end{pmatrix},
\end{equation}
which, expressed in terms of $\alpha$, takes the form
\begin{equation}
Z_R = \begin{pmatrix} -{\rm sin}\alpha & {\rm cos}\alpha \\
{\rm cos}\alpha  & {\rm sin}\alpha \end{pmatrix}.
\end{equation}
The $\alpha$, at tree level, can be expressed as
\begin{equation}
{\rm tan}\alpha = \frac{-(M^2_{A^0}-M^2_Z){\rm cos}2\beta -
\sqrt{(M^2_{A^0}+M^2_Z)^2-4M^2_{A^0} M^2_Z {\rm
cos}^22\beta}}{(M^2_{A^0}+M^2_Z){\rm sin}2\beta}.
\label{eq:dependence}
\end{equation}
Notice that $\alpha$ is approximately $\beta-\pi/2$ at the limit
$M_{A^0}>>M_Z$. Radiative corrections modify the latter relation,
and we include these effects in our numerical study.

The loop function $j(r,r')$ is the same as in Ref.
\cite{Giudice:2005rz}
\begin{equation}
j(r)=\frac{r {\rm Log}r}{r-1}, ~~~ j(r,r')=\frac{j(r)-j(r')}{r-r'}.
\end{equation}

\end{document}